\documentclass[twoside]{ilcws08}
\usepackage[latin1]{inputenc}
\usepackage[dvips]{graphicx,epsfig,color}
\usepackage{wrapfig,rotating}
\usepackage{amssymb,amsmath,array}

\begin{document}

\title{
Reduction Method for One-loop Tensor 5- and 6-point Integrals Revisited}
%***********************************************************************
% AUTHORS INFORMATION AREA
%***********************************************************************
\author{Theodoros Diakonidis
% Optional short acknowledgment: remove next line if non-needed
%\thanks{This is an optional funding source acknowledgment.}
% DO NOT MODIFY THE FOLLOWING '\vspace' ARGUMENT
\vspace{.3cm}\\
% Addresses and institutions (remove "1- " in case of a single institution)
~Deutsches Elektronen-Synchrotron, DESY, Platanenallee
  6, 15738 Zeuthen, Germany \\
}
%%***********************************************************************
% END OF AUTHORS INFORMATION AREA
%***********************************************************************

\maketitle

\begin{abstract}
A complete analytical reduction of general one-loop Feynman
integrals with five legs for tensors up to rank $R=3$
and six legs for tensors up to rank $4$ is reviewed \cite{chicagopresent:2008}.
An elegant formalism with extensive use of signed
minors was developed for the cancellation of leading inverse Gram determinants.
The resulting compact
formulae allow both for a study of analytical properties and for efficient
numerical programming. Here some special numerical examples are presented.
\end{abstract}

\section{Introduction}

The Feynman integrals for reactions with up to four external particles
have been systematically studied and evaluated in numerous studies.
It is needed to be mentioned here the seminal papers \cite{'tHooft:1979xw} and
\cite{Passarino:1979jh}
and the Fortran packages \texttt{FF} \cite{vanOldenborgh:1990yc} and
\texttt{LoopTools} \cite{Hahn:1998yk}, which evidently show the situation so far.
The treatment of Feynman integrals with a higher multiplicity than
four becomes quite involved if questions of efficiency and stability
become vital, as it happens with the calculational problems related to
high-dimensional phase space integrals over sums of thousands of Feynman
diagrams with internal loops.

What is reviewed here is an approach which reduces the tensor
integrals algebraically to sums over a small set of scalar two-, three-
and four-point functions, which are assumed to be known.
To accomplish this methods
of ref.~\cite{Fleischer:1999hq,Davydychev:1991va} are used.
The present goal is to provide compact analytic formulas for the
complete reduction of tensor pentagons and hexagons to scalar
master integrals, which are free of leading inverse Gram determinants.
For a study of gauge invariance and of the ultraviolet
(UV) and infrared (IR) singularity structure of a set of Feynman diagrams,
it is evident that a complete reduction  is advantageous, and it may
also be quite useful for a tuned, analytical study of certain regions
of potential numerical instabilities.

The numerics are obtained with two independent implementations, one made
in Mathematica, and another one in Fortran.
The Mathematica program \texttt{hexagon.m} with the reduction formulae
is made publicly available
\cite{hexagon:2008}, see also \cite{Diakonidis:2008dt} for a short description.
For numerical applications, one has to link the package with a program for the
evaluation of scalar one- to four-point functions, e.g. with
\texttt{LoopTools} \cite{Hahn:1998yk,Hahn:2006qw,vanOldenborgh:1990yc},
\texttt{CutTools} \cite{vanHameren:2005ed,Ossola:2007ax},
\texttt{QCDLoop} \cite{Ellis:2007qk}.
\section{Useful Notations}
It is useful to introduce the notation for the loop integrals
and also for certain determinants
that occur in the recurrence relations and their solutions.
The one-loop, $N$-point tensor integrals of rank $R$
in $d$-dimensional space-time are defined as,
\begin{equation}
\label{eq:JNR}
I^{(N)}_{\mu_1\ldots\mu_R} \left(d;\nu_1,\ldots,\nu_N\right) =
\int \frac{d^d k}{i \pi^{d/2}}
\frac{k_{\mu_1}\ldots k_{\mu_R}}{D_1^{\nu_1}\ldots D_N^{\nu_N}}
% \, ,
\end{equation}
with propagator denominators
\begin{equation}
% D_j = (k+q_j)^2 - m_j^2 + i \epsilon \, .
D_j = (k-q_j)^2 - m_j^2 + i \epsilon \, .
\end{equation}
The determinant of an $(N+1)\times(N+1)$ matrix, known as the modified
Cayley determinant is defined as:~\cite{Melrose:1965kb},
\begin{equation}
\label{eq:smN}
()_N ~\equiv~  \left|
\begin{array}{ccccc}
  0 & 1       & 1       &\ldots & 1      \\
  1 & Y_{11}  & Y_{12}  &\ldots & Y_{1N} \\
  1 & Y_{12}  & Y_{22}  &\ldots & Y_{2N} \\
  \vdots  & \vdots  & \vdots  &\ddots & \vdots \\
  1 & Y_{1N}  & Y_{2N}  &\ldots & Y_{NN}
\end{array}
\right| \, ,
\end{equation}
with coefficients
\begin{equation}
Y_{ij}=-(q_i-q_j)^2+m_i^2+m_j^2 \, , \quad (i,j = 1 \ldots N) \, .
\end{equation}

All other determinants appearing are signed minors of $()_N$, constructed
by deleting $m$ rows and $m$ columns
from $()_N$, and multiplying with a sign factor. They will be denoted
by
\begin{eqnarray}
\lefteqn{
\left(
\begin{array}{cccc}
  j_1 & j_2 & \cdots & j_m\\
  k_1 & k_2 & \cdots & k_m\\
\end{array}
\right)_N
~\equiv~ {(-1)}^{\sum_l (j_l + k_l)}
} && \nonumber \\ &&
% \times
\hspace{-1.5em}
% \sigma_{\{j\}} \, \sigma_{\{k\}} \,
\mbox{sgn}_{\{j\}} \, \mbox{sgn}_{\{k\}} \,
 \left|
\begin{array}{c}
\mbox{rows $j_1\cdots j_m$ deleted}\\
\mbox{columns $k_1\cdots k_m$ deleted}\\
\end{array}
 \right| \, ,
\end{eqnarray}
where $\mbox{sgn}_{\{j\}}$ and $\mbox{sgn}_{\{k\}}$ are the signs of
permutations that sort the deleted rows $j_1\cdots j_m$ and columns
$k_1\cdots k_m$ into ascending order.

\section{Pentagons}
In this chapter final results are provided
of the reduction concerning ranks up to 3.
More about these can be found in \cite{Diakonidis:2008ij}.
For the {\bf scalar} 5-point function the recursion relation
for the limit of $d=4$ is,
\begin{eqnarray}\label{i5sc2}
E \equiv I_{5}
&=&
\frac{1}
{
           \begin{pmatrix} 0 \\ 0\end{pmatrix}_5
}
\sum_{s=1}^{5} {\begin{pmatrix} 0 \\ s \end{pmatrix} }_5
I_{4}^{s},
\label{scalar4p}
\end{eqnarray}

Similarly, for the tensor integral
of {\bf rank 1} (vector) in the limit $d \to 4$ we obtain:
\begin{eqnarray}\label{i5vc1}
 I_{5}^{\mu}
&=&
\sum_{i=1}^{4} \, q_i^{\mu} I_{5,i},
\\
I_{5,i} \equiv E_i
%\nl
&=&
-
\frac{1}{
\begin{pmatrix} 0 \\ 0 \end{pmatrix}_5
}
\sum_{s=1}^{5} \begin{pmatrix} 0&i \\ 0&s \end{pmatrix}_5
I_{4}^{s},
\label{first}
\end{eqnarray}

The tensor integral of {\bf rank 2} can be written:
\begin{eqnarray}
I_{5}^{\mu\, \nu\,}&=& \sum_{i,j=1}^{4} \, q_i^{\mu}\, q_j^{\nu} E_{ij} +
g^{\mu \nu}  E_{00},
\\
E_{ij}&=&\sum_{s=1}^{5} S_{ij}^{4,s} I_4^s +\sum_{s,t=1}^{5} S_{ij}^{3,st} I_3^{st} ,
\label{final2}
\\
E_{00}&=&-\frac{1}{2}
\frac{1}{
{\begin{pmatrix} 0 \\ 0 \end{pmatrix} }_5
} \sum_{s=1}^5
\frac{
{\begin{pmatrix} s \\ 0 \end{pmatrix} }_5
}
{
{\begin{pmatrix} s \\ s \end{pmatrix} }_5
}
\left[
{\begin{pmatrix} 0&  s \\ 0&s \end{pmatrix} }_5
I_4^s
-
\sum_{t=1}^{5}
{\begin{pmatrix} t&  s \\ 0&s \end{pmatrix} }_5
I_3^{st} \right] \, .
\end{eqnarray}
Finally the tensor integral of {\bf rank 3}
\begin{eqnarray}
\label{final3b}
%\nonumber
I_{5}^{\mu\, \nu\, \lambda}
&=&
\sum_{i,j,k=1}^{4} \, q_i^{\mu}\, q_j^{\nu} \, q_k^{\lambda}
E_{ijk}+\sum_{k=1}^4 g^{[\mu \nu} q_k^{\lambda]} E_{00k},
\\
%\nonumber
E_{ijk}&=&\sum_{s=1}^{5} S_{ijk}^{4,s} I_4^s +\sum_{s,t=1}^{5} S_{ijk}^{3,st} I_3^{st}+
\sum_{s,t,u=1}^{5} S_{ijk}^{2,stu} I_2^{stu} ,
\label{final3a}
\\
E_{00k}&=&\sum_{s=1}^{5} S_{00k}^{4,s} I_4^s +\sum_{s,t=1}^{5} S_{00k}^{3,st} I_3^{st}+
\sum_{s,t,u=1}^{5} S_{00k}^{2,stu} I_2^{stu} .
\label{final3c}
\end{eqnarray}
All coefficients,also those not explicitly defined here $S_{ijk}^{4,s}, S_{ijk}^{3,st}, S_{ijk}^{2,stu}$,
$S_{00k}^{4,s}$, $S_{00k}^{3,st}$, $S_{00k}^{2,stu}$, $S_{ij}^{4,s}$, $S_{ij}^{3,st}$ (see \cite{Diakonidis:2008ij}),
are free of the leading Gram determinants.
\section{Hexagons}
If the external momenta of a hexagon are 4-dimensional,
their Gram determinant vanishes: $ \left( \right)_6 = 0 $, and
a linear relation between the propagators $D_j$ exists:
\begin{equation}
1 ~=~
\sum_{j=1}^6 \frac{{\begin{pmatrix} 0 \\ j \end{pmatrix} }_6} {{\begin{pmatrix} 0 \\ 0 \end{pmatrix} }_6} D_j \, .
\end{equation}
With this relation, any hexagon integral can trivially be
reduced to pentagons. For example, for the scalar hexagon,
one obtains the well-known result~\cite{Melrose:1965kb}:
\begin{equation}
I_6=\sum_{r=1}^6 \frac{{\begin{pmatrix} 0 \\ r \end{pmatrix}_6 }}{\begin{pmatrix} 0 \\ 0 \end{pmatrix} }_6 I_5^r \, ,
\end{equation}
where the scalar pentagon $I_5^r$ on the right hand side is
obtained by removing line $r$ from the hexagon $I_6$.
In the same way, tensor hexagons of rank $R$ can be reduced to tensor
pentagons of rank $R$. However, it was noticed in
ref.~\cite{Fleischer:1999hq} that a reduction directly to tensor
pentagons of rank $R-1$ is also possible:
\begin{equation}
\label{eq:I6tensor}
I_6^{\mu_1\ldots\mu_R} ~=~
 \sum_{r=1}^6 v_r^{\mu_1} I_5^{\mu_2\ldots\mu_R\,,r} \, ,
\end{equation}
where
\begin{equation}
v_r^{\mu} ~\equiv~
 - \frac{1}{{\begin{pmatrix} 0 \\ 0 \end{pmatrix} }_6} \sum_{i=1}^{5} {{\begin{pmatrix} 0& i \\ 0& r \end{pmatrix} }_6} q_i^{\mu} \, .
\end{equation}
A more general proof of this property was given in ref.~\cite{Denner:2005nn}.
By substituting the reduction formulas for tensor pentagons into
eq.~(\ref{eq:I6tensor}), we can immediately express tensor
hexagons in terms of scalar master integrals.
In this way using the formulas of the previous section we can provide results for integrals
up to 4th rank for the hexagons (see \cite{Diakonidis:2008ij}).
\section{Numerical results}
\begin{figure}[t]
\begin{center}
\includegraphics[scale=0.90]{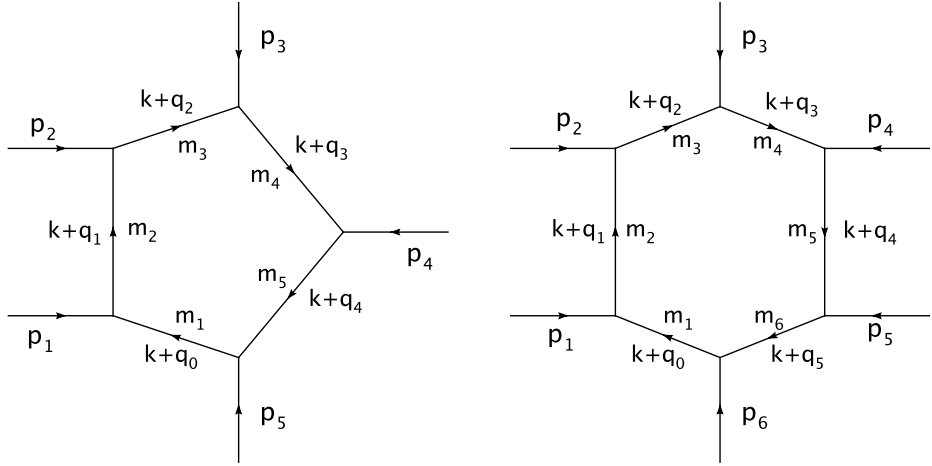}
\caption{
Momenta flow used in the numerical examples for six- and five-point integrals.}
\label{fig:6pt5ptfig}
\end{center}
\end{figure}

%% section headers !
In order to illustrate the numerical results which can be obtained with
the described approach, a representative collection of
tensor components will be evaluated, for some special cases which are not included in \cite{Diakonidis:2008dt}.
The kinematics are visualized in Figure ~\ref{fig:6pt5ptfig}.
%------------------------------------------------------------------------------
%------------------------------------------------------------------------------

For the evaluation of the scalar two-, three- and four-point functions,
which appear after the complete reduction,
we have implemented two numerical libraries:
\begin{itemize}
\item
For massive internal particles:
\texttt{Looptools 2.2} \cite{Hahn:1998yk,Hahn:2006qw};
\item
If there are also massless internal particles:
\texttt{QCDLoop-1.4} \cite{Ellis:2007qk}.
\end{itemize}

the first one in the published package \texttt{hexagon.m} \cite{hexagon:2008} and both of them in the Fortran implimentation.
%---------------------------------------------------------------------
\begin{table}[htb]%third - massless - phase space point
\centering
\begin{tabular}{|r|r@{.}l|r@{.}l|r@{.}l|r@{.}l|}
\hline

$p_1$ &   5&0 &   0&0 &   0&0 &   5&0 \\
$p_2$ &   5&0 &   0&0 &   0&0 &  -- 5&0 \\
$p_3$ & -- 1&6554963633 &   1&2970338732 &  -- 0&9062452085 &  -- 0&4869198730 \\
$p_4$ & -- 3&8970139847  &  0&0528728505  & -- 2&5360890226  &  2&9584074987 \\
$p_5$ & -- 4&4474896520  & -- 1&3499067237  &  3&4423342311  & -- 2&4714876256  
\\
\hline
\multicolumn{9}{|c|}{$m_1 = \cdots = m_5 = 0.0$}
\\
\hline
\end{tabular}
\caption{\label{massless-phasespacepoint}
The external four-momenta  for the five-point functions;
all internal and external particles are massless.}
\end{table}
\begin{table}
 \centering
\begin{tabular}{|l| r@{ i }l | r@{ i }l |}
\hline
&\multicolumn{2}{|c|}{$\epsilon^0$}
&\multicolumn{2}{|c|}{$1/\epsilon$}
\\
\hline
$E_0$& 0.49975096E-03 +&0.12807271E-02 &0.33696138E-03 --&0.64416161E-03\\
\hline
$E^{1}$&-- 0.50336057E-03 --&0.10928553E-02&-- 0.34786666E-03 +&0.54767334E-03\\
\hline
$E^{12}$& -- 0.11603164E-02 --&0.17552616E-02 &-- 0.60899168E-03 +&0.12327007E-02\\
\hline
$E^{122}$&-- 0.43997517E-02 --&0.34454891E-02 &-- 0.10597882E-02 +&0.36519758E-02\\
\hline
\end{tabular}
\vspace{.1cm}\\
\begin{tabular}{|l| r@{  i }l |}
\hline
&\multicolumn{2}{|c|}{$1/\epsilon^2$}\\
\hline
$E_0$   &-- 0.15779987E-03+&0.00000000E+00   \\
\hline
$E^{1}$ & 0.17432984E-03+&0.00000000E+00    \\
\hline
$E^{12}$&   0.39238082E-03+& 0.00000000E+00 \\
\hline
$E^{122}$& 0.11624600E-02+& 0.00000000E+00  \\
\hline
\end{tabular}

\caption[]{\label{5-point-massless}
Selected tensor components of five-point tensor functions with massless particles; kinematics defined in
Table \ref{massless-phasespacepoint} (Cross checked with \texttt{golem95} \cite{Binoth:2008uq}).}
\end{table}

%---------------------------------------------------------------------

%=========================================================================
\begin{table}[b]%second phase space point

\centering
\begin{tabular}{|r|r@{.}l|r@{.}l|r@{.}l|r@{.}l|}
\hline
$p_1$ & 5&0 & 0&0 & 0&0 & 5&0 \\
$p_2$ & 5&0 & 0&0 & 0&0 &-- 5&0 \\
$p_3$ & -- 0&7623942818&  0&5390582570&  -- 0&5220507689 &   0&1346262645 \\
$p_4$ &-- 3&3298826057 & -- 1&0349623069 & -- 1&1048040197 &   2&9658690580 \\
$p_5$ & -- 2&8267285956 & -- 1&4136906402  &  2&3189438782 &  -- 0&7838192500 \\
$p_6$ & -- 3&0809945169 &  1&9095946901&  -- 0&6920890895 &  -- 2&3166760725 
\\
\hline
\multicolumn{9}{|c|}{$m_1 = 1.0, \;
m_2 = 1.2, ~~\;
m_3 = 1.4, ~~\;
m_4 = 1.6, ~~\;
m_5 = 1.8, ~~\;
m_6 = 2.0$}
\\
\hline
\end{tabular}
\caption{\label{phasespacepoint-1}
The external four-momenta  for the six-point functions; all external legs massless
and the internal massive.}
\end{table}

\begin{table}[htb]
\centering
\begin{tabular}{|l| r@{ i }l|}
\hline
$F_0$ &   0.54701021E-04   -- &0.67031213E-04     \\
\hline
$F^{3}$&  -- 0.32082506E-04   + &0.24545301E-03   \\
\hline
$F^{11}$& 0.13862332E-04   -- &0.12247788E-03    \\
\hline
$F^{112}$&-- 0.22452724E-04 -- &0.39826579E-04  \\
\hline
$F^{0121}$& 0.15817785E-03  + &0.26882173E-03 \\
\hline
\end{tabular}

\caption[]{\label{6-point-massive}
Selected tensor components of six-point tensor functions produced by the phase space point of
Table \ref{phasespacepoint-1}.}
\end{table}

\clearpage
\section{Acknowledgments}

Work supported the European Community's Marie-Curie Research Trai\-ning Networks
MRTN-CT-2006-035505 ``HEPTOOLS'' and by Sonderforschungsbereich/Transregio SFB/TRR 9 of DFG ``Com\-pu\-ter\-ge\-st\"utz\-te Theoretische Teilchenphysik".
I would also like to thank my collaborators  J. Fleischer,
J. Gluza, K. Kajda, T. Riemann,
and especially J. B. Tausk
for useful discussions.

\begin{footnotesize}

% IF YOU USE BIBTEX,
% - DELETE THE TEXT BETWEEN THE TWO ABOVE DASHED LINES
% - UNCOMMENT THE NEXT TWO LINES AND REPLACE 'Name_Of_Your_BibFile'

\bibliographystyle{unsrt}
\bibliography{2loops_teo}
% example of Name_Of_Your_BibFile.bib
%@misc{chicagopresent:2008,
%   note =        "Presentation \newline
%                  http://ilcagenda.linearcollider.org/materialDisplay.py?contribId=77&sessionId=18&materialId=slides&confId=2628"
%}
\end{footnotesize}

% ****************************************************************************
% END OF BIBLIOGRAPHY AREA
% ****************************************************************************

\end{document}